\documentclass[11pt]{article}
\usepackage{amsmath}
\usepackage{amssymb}
\usepackage[dvips]{epsfig}

 \advance \textwidth by -2in
 \advance \textheight by -3in
\def\##1{{\underline #1}}
\def\*#1{\tilde{{\underline #1}}}
\def\=#1{\underline{\underline{#1}}}
\def\+#1{\tilde{\underline{\underline{#1}}}}

\def\le{\left(}
\def\ri{\right)}
\def\les{\left[}
\def\ris{\right]}

\def\c#1{\cite{#1}}

\def\r#1{(\ref{#1})}

\def\.{\mbox{ \tiny{$^\bullet$} }}
\def\,{\thinspace}

\def\rt{\le \#x,t\ri}
\def\ro{\le \#x,\omega\ri}
\def\partialt{\frac{\partial}{\partial t}\,}
\def\curl{\nabla\times}
\def\div{\nabla\.}

\def\epso{\epsilon_{\scriptscriptstyle 0}}

\def\muo{\mu_{\scriptscriptstyle 0}}

\def\eps{\epsilon}

\begin{document}

\vskip 0.4cm

\begin{center}
{\large {\bf On the genesis of Post constraint in modern electromagnetism}}
\vskip 0.2cm

\noindent  {Akhlesh Lakhtakia}\footnote{Tel: +1 814 863 4319; Fax: +1 814 865 9974;
E--mail: akhlesh@psu.edu}
\vskip 0.2cm
\noindent {\em Computational \& Theoretical Materials Sciences Group (CATMAS)\\
Department of Engineering Science \& Mechanics\\
Pennsylvania State University, University Park, PA 16802--6812, USA}

\end{center}

\noindent {\bf Abstract:} The genesis of the Post constraint is premised on two attributes of modern
electromagnetism: (i) its microscopic nature, and (ii) the status of
$\*e\rt$ and $\*b\rt$ as the primitive electromagnetic fields. This constraint can therefore not arise 
in EH--electromagnetism, wherein the primitive electromagnetic fields are the macroscopic
fields $\*E\rt$ and $\*H\rt$. Available experimental evidence against the Post constraint is incomplete
and inconclusive.


\section{Introduction}
Ever since its enunciation in 1962 \c{Post}, the Post constraint has been an enigma. It was ignored for over three decades
by the electromagnetics community for reasons that will probably be extracted only by future
historians of science. It arose from obscurity like a phoenix in 1994 in the context of linear, nonreciprocal, biisotropic mediums \c{LWmtt},
and since then has been the subject of discussion in the complex--mediums electromagnetics research community.

A remarkable feature of the Post constraint is that it permits a sharp distinction between two widely prevalent conceptions of
electromagnetic phenomenons. The genesis of the Post constraint lies in the microphysical basis of
{\em modern electromagnetism}, whereby the (necessarily macroscopic) constitutive functions must be conceived as
piecewise homogeneous entities and can therefore not vary continuously in spacetime. In contrast, {\em EH--electromagnetism}
is essentially macroscopic, and its principles seem to be inimical to the validity of the Post constraint. Available
experimental evidence does not negate the Post constraint, but cannot be held to be conclusive either.

These issues are discussed in this essay. Section \ref{ModEMtheo} is an exposition of modern electromagnetism encompassing 
both the microscopic and the macroscopic levels. Section \ref{PostCons} presents the rationale for and the genesis
of the Post constraint. The characteristics of EH--electromagnetism relevant to the Post constraint are given in Section \ref{EHEM},
while experimental evidence is reviewed in Section \ref{ExpEvi}. Finally, in Section \ref{PCfs} the constitutive equations
of free space are deduced in relation to the Post constraint.

\section{Modern Electromagnetism}
\label{ModEMtheo}
Electromagnetism today is a microscopic science, even though it is
mostly used in its macroscopic form. It was certainly a macroscopic
science when Maxwell 
unified the equations of Coulomb, 
Gauss, 
Faraday, 
and Amp\`ere, 
added a displacement current to Amp\`ere's
equation, and produced  the four equations to which his name
is attached.
Although Maxwell had abandoned a mechanical basis for 
electromagnetism during the  early 1860s, and even used terms like
{\em molecular vortices\/}, a close reading \c{TKS97}
of his papers will
convince the reader that Maxwell's conception of electromagnetism~---~like that of most of
his contemporaries~---~was
macroscopic. 

By the end of the 19th century, that conception had been
drastically altered \c{JZB85}. Hall's 
successful explanation of the
eponymous effect, the postulation of the electron by Stoney
and its subsequent discovery by Thomson, 
and Larmor's 
theory of the electron precipitated that
alteration. It was soon codified by Lorentz  and
Heaviside,  so that the
20th century dawned with the acquisition of a microphysical basis 
by electromagnetism. Maxwell's
equations remained unaltered in form at macroscopic
length scales, but their roots now lie in the fields engendered
by microscopic charge quantums. The subsequent emergence
of {\em quantum mechanics\/} 
did not change the form
of the macroscopic equations either,
although the notion of a field lost its determinism
and an inherent uncertainty was recognized in the measurements
of key variables \c{Sch58}.

\subsection{Microscopic Maxwell Postulates}
The microscopic fields
 are just two: the
electric field $\tilde{\#e}\rt$
and the magnetic field $\tilde{\#b}\rt$.\footnote{The lower--case 
letter signifies that
the quantity is microscopic,
while the tilde $\tilde{ }$
indicates dependence on time.} These two are accorded the status
of primitive fields in modern electromagnetism. 
Both fields vary extremely rapidly as
functions of position $\#x$ and time $t$.
Their sources are the microscopic charge
density ${\tilde c}\rt$ 
and the microscopic current density
$\tilde{\#j}\rt$, where
\begin{eqnarray}
\label{eq6.1}
{\tilde c}\rt = \sum_\ell\,q_\ell\,\delta\les\#x-\#x_\ell(t)\ris\,,
\\
\label{eq6.2}
\tilde{\#j}\rt= \sum_\ell\,q_\ell\#v_\ell\,\delta\les\#x-
\#x_\ell(t)\ris\,;
\end{eqnarray}
$\delta(\.)$ is the Dirac delta function; while
$\#x_\ell(t)$ and $\#v_\ell(t)$ are the position and
the velocity of the point charge $q_\ell$.
Uncertainties in the measurements of the positions and the 
velocities
of the discrete point charges open the door to
quantum mechanics,
but we need
not traverse that path here.

All of the foregoing fields and sources appear in the {\em 
microscopic\/}
Maxwell postulates:
\begin{eqnarray}
\label{eq6.3}
&&\div \tilde{\#e}\rt = \epso^{-1}\,\tilde{c}\rt\,,
\\
\label{eq6.4}
&&\curl\tilde{\#b}\rt -\epso\muo\partialt\tilde{\#e}\rt = 
\muo\,\tilde{\#j}\rt\,,
\\
\label{eq6.5}
&&\div\tilde{\#b}\rt = 0\,,
\\
\label{eq6.6}
&&
\curl\tilde{\#e}\rt +
\partialt\tilde{\#b}\rt = \#0\,.
\end{eqnarray}
In these equations and hereafter, $\epso = 8.854\times 10^{-12}$~F/m and
$\muo = 4\pi\times 10^{-7}$~H/m are the permittivity 
\index{permittivity}
and the
permeability \index{permeability}
of free space (i.e., vacuum), respectively.
The first two postulates are inhomogeneous differential
equations as they contain source terms on their right sides,
while the last two are homogeneous differential equations.

\subsection{Macroscopic Maxwell Postulates}
Macroscopic measuring devices average over (relatively) large 
spatial and
temporal intervals. Therefore, spatiotemporal averaging of the
microscopic quantities appears necessary in order to deduce the
{\em macroscopic} Maxwell postulates
from \r{eq6.3}--\r{eq6.6}. Actually, only spatial averaging is 
necessary
\c{Jack},
because it implies temporal averaging due to the finite magnitude
of the universal maximum speed 
$\le\epso\muo\ri^{-1/2}$.
Denoting  the macroscopic charge
and current densities, respectively, by $\tilde\rho\rt$
and $\tilde{\#J}\rt$, we obtain the {\em macroscopic\/}
Maxwell
postulates 
\index{Maxwell postulates!macroscopic}
\begin{eqnarray}
\label{eq6.7}
&&\div \tilde{\#E}\rt = \epso^{-1}\,\tilde{\rho}\rt\,,
\\
\label{eq6.8}
&&\curl\tilde{\#B}\rt -\epso\muo\partialt\tilde{\#E}\rt = 
\muo\,\tilde{\#J}\rt\,,
\\
\label{eq6.9}
&&\div\tilde{\#B}\rt = 0\,,
\\
\label{eq6.10}
&&
\curl\tilde{\#E}\rt +
\partialt\tilde{\#B}\rt = \#0\,,
\end{eqnarray}
which involve the macroscopic primitive fields
\index{macroscopic field}\index{field!macroscopic}
$\tilde{\#E}\rt$ and $\tilde{\#B}\rt$
as the spatial averages of $\tilde{\#e}\rt$ and $\tilde{\#b}\rt$,
respectively.
From \r{eq6.7} and \r{eq6.8}, a macroscopic continuity
equation  for the source densities 
can be derived as 
\begin{equation}
\label{eq6.11}
\div\tilde{\#J}\rt +\partialt\tilde\rho\rt=0\,.
\end{equation}

\subsection{Familiar Form of Macroscopic Maxwell Postulates}
Equations \r{eq6.7}--\r{eq6.10} are not the familiar form
of the macroscopic Maxwell postulates, even though they hold 
in free space as well as in matter. The familiar form emerges
after the recognition that matter contains, in general, both free
charges and bound charges. Free and bound source densities can be 
decomposed as
\begin{equation}
\label{eq6.12}
\tilde\rho\rt = {\tilde\rho}_{so}\rt - \div\tilde{\#P}\rt\,
\end{equation}
and
\begin{equation}
\label{eq6.13}
\tilde{\#J}\rt = {\tilde{\#J}}_{so}\rt + \partialt\tilde{\#P}\rt
+\curl\tilde{\#M}\rt\,.
\end{equation}
This decomposition is consistent with \r{eq6.11}, provided
the free source densities obey the reduced continuity equation
\index{continuity equation, macroscopic!reduced}
\begin{equation}
\label{eq6.14}
\div\tilde{\#J}_{so}\rt +\partialt{\tilde\rho}_{so}\rt=0\,.
\end{equation}
The free source densities represent ``true" sources which can be
externally impressed. Whereas $\tilde{\#J}_{so}\rt$ is the
source current density, ${\tilde\rho}_{so}\rt$ is the
source charge density.

Bound source densities represent matter in its macroscopic form
and are, in turn, quantified by the polarization ${\tilde{\#P}}\rt$ and the {magnetization} ${\tilde{\#M}}\rt$. Both ${\tilde{\#P}}\rt$ and
${\tilde{\#M}}\rt$ are nonunique to the extent that they
can be replaced by ${\tilde{\#P}}\rt-\curl {\tilde{\#A}}\rt$
and  ${\tilde{\#M}}\rt+(\partial/\partial t)\,{\tilde{\#A}}\rt$,
respectively, in \r{eq6.12} and \r{eq6.13} without affecting
the left sides of either equation.

Polarization and magnetization are subsumed in the definitions
of the electric induction $\tilde{\#D}\rt$ and the magnetic 
induction $\tilde{\#H}\rt$ as follows:
\begin{eqnarray}
\label{eq6.15}
&&\tilde{\#D}\rt = \epso\,\tilde{\#E}\rt + \tilde{\#P}\rt\,,
\\
\label{eq6.16}
&&\tilde{\#H}\rt = \muo^{-1}\,\tilde{\#B}\rt - \tilde{\#M}\rt\,.
\end{eqnarray}
Then, \r{eq6.7}--\r{eq6.10} metamorphose into the
familiar form of the  macroscopic Maxwell postulates:
\begin{eqnarray}
\label{eq6.17}
&&\div \tilde{\#D}\rt = \tilde{\rho}_{so}\rt\,,
\\
\label{eq6.18}
&&\curl\tilde{\#H}\rt -\partialt\tilde{\#D}\rt = 
\tilde{\#J}_{so}\rt\,,
\\
\label{eq6.19}
&&\div\tilde{\#B}\rt = 0\,,
\\
\label{eq6.20}
&&
\curl\tilde{\#E}\rt +
\partialt\tilde{\#B}\rt = \#0\,.
\end{eqnarray}

Let us note, in passing, that the fields $\tilde{\#d}\rt$ and $\tilde{\#h}\rt$
do not exist in microphysics, matter being an
ensemble of point charges in free space. 

\subsection{Linear Constitutive Relations}

The induction fields at some point in spacetime
$({\#x},t)$ can depend {\em locally} on
the primitive fields at the same $({\#x},t)$. This dependence
can be spatially
nonhomogeneous (i.e., dependent on space ${\#x}$) and/or can vary with
time $t$ (i.e., age). In addition, the induction fields at $({\#x},t)$ can depend
nonlocally on the primitive fields at some $({\#x}-{\#x}_h,t-t_h)$,
where the spacetime interval $({\#x}_h,t_h)$, $t_h \geq 0$, must be timelike
in order to be causally influential \cite[pp. 85--89]{LucH}. Thus, the most general
linear constitutive relations \c{LW98}
\begin{eqnarray}
\nonumber
&&\*D\rt = \int\int \  \+\eps({\#x},t;{\#x}_h,t_h)\.\*E({\#x}-{\#x}_h,t-t_h) \,
d{\#x}_h\, dt_h
\\[5pt]
\label{dconrel1}
& &\qquad+\int\int \  \+\xi({\#x},t;{\#x}_h,t_h)\.\*B({\#x}-{\#x}_h,t-t_h) \,
d{\#x}_h \,dt_h \, 
\end{eqnarray}
and
\begin{eqnarray}
\nonumber
&&\*H\rt = \int\int \  \+\zeta({\#x},t;{\#x}_h,t_h)\.\*E({\#x}-{\#x}_h,t-t_h) \,
d{\#x}_h\, dt_h
\\[5pt]
\label{hconrel1}
&&\qquad+\int\int \  \+\nu({\#x},t;{\#x}_h,t_h)\.\*B({\#x}-{\#x}_h,t-t_h) \,
d{\#x}_h\, dt_h \, 
\end{eqnarray}
can describe any linear medium~---~indeed, the entire universe after linearization.
The integrals extend only over the causal values of
$({\#x}_h,t_h)$, but that does not restrict the analysis presented here.

\section{The Post Constraint}
\label{PostCons}

Four second--rank tensors appear in the foregoing constitutive relations:
$\+\eps$ is the permittivity tensor,
$\+\nu$ is the impermeability tensor,
while $\+\xi$  and
$\+\zeta$ are
the magnetoelectric tensors. Together, these four tensors contain 36 scalar
functions; but the Post constraint indicates that only 35, at most, are independent.
This was clarified elsewhere \c{LW96} using 4--tensor notation, but we
revisit the issue here for completeness.
Let us therefore express the magnetoelectric tensors as
\begin{equation}
\+\xi({\#x},t;{\#x}_h,t_h) =\+\alpha({\#x},t;{\#x}_h,t_h) +\frac{1}{6}\,\=I\,{\tilde\Psi}({\#x},t;{\#x}_h,t_h)\,
\end{equation}
and
\begin{equation}
\+\zeta({\#x},t;{\#x}_h,t_h) =\+\beta({\#x},t;{\#x}_h,t_h) -\frac{1}{6}\,\=I\,{\tilde\Psi}({\#x},t;{\#x}_h,t_h)\,,
\end{equation}
where $\=I$ is the identity tensor and the scalar function
\begin{equation}
{\tilde\Psi}({\#x},t;{\#x}_h,t_h) = {\rm Trace}\Big(\+\xi({\#x},t;{\#x}_h,t_h) -\+\zeta({\#x},t;{\#x}_h,t_h) \Big)\,.
\end{equation}
Therefore,
\begin{equation}
{\rm Trace}\Big(\+\alpha({\#x},t;{\#x}_h,t_h) -\+\beta({\#x},t;{\#x}_h,t_h) \Big) \equiv 0\,.
\end{equation}

\subsection{Rationale for the Post Constraint}
Let us recall that \r{eq6.19} and \r{eq6.20} do not contain the induction fields $\*D\rt$ and $\*H\rt$. Hence, \r{dconrel1} and \r{hconrel1}
must be substituted only in \r{eq6.17} and \r{eq6.18};
thus,
\begin{eqnarray}
\nonumber
&&
\int\int \  \div\Big(\+\eps({\#x},t;{\#x}_h,t_h)\.\*E({\#x}-{\#x}_h,t-t_h)
\\
\nonumber
&&\qquad\qquad+\+\alpha({\#x},t;{\#x}_h,t_h)\.\*B({\#x}-{\#x}_h,t-t_h)\Big) \,
d{\#x}_h\, dt_h
\\
\nonumber
&&+\frac{1}{6}\,\int\int \  {\tilde\Psi}({\#x},t;{\#x}_h,t_h) \Big(\div\*B({\#x}-{\#x}_h,t-t_h) \Big)\,
d{\#x}_h\, dt_h
\\
\nonumber
&&+\frac{1}{6}\,\int\int \  \Big(\nabla{\tilde\Psi}({\#x},t;{\#x}_h,t_h)\Big) \.\*B({\#x}-{\#x}_h,t-t_h) \,
d{\#x}_h\, dt_h
\\
&&\qquad\qquad\quad= \tilde{\rho}_{so}\rt\,
\label{eq6.17a}
\end{eqnarray}
and
\begin{eqnarray}
\nonumber
&&
\int\int \  \curl\Big(\+\beta({\#x},t;{\#x}_h,t_h)\.\*E({\#x}-{\#x}_h,t-t_h)
\\
\nonumber
&&\qquad\qquad+\+\nu({\#x},t;{\#x}_h,t_h)\.\*B({\#x}-{\#x}_h,t-t_h)\Big) \,
d{\#x}_h\, dt_h
\\
\nonumber
&&
-\int\int \  \partialt\Big(\+\eps({\#x},t;{\#x}_h,t_h)\.\*E({\#x}-{\#x}_h,t-t_h)
\\
\nonumber
&&\qquad\qquad+\+\alpha({\#x},t;{\#x}_h,t_h)\.\*B({\#x}-{\#x}_h,t-t_h)\Big) \,
d{\#x}_h\, dt_h
\\
\nonumber
&&-\frac{1}{6}\,\int\int \  {\tilde\Psi}({\#x},t;{\#x}_h,t_h) \Big(\curl\*E({\#x}-{\#x}_h,t-t_h) 
\\
\nonumber
&&
\qquad\qquad+
\partialt\*B({\#x}-{\#x}_h,t-t_h) \Big)\,
d{\#x}_h\, dt_h
\\
\nonumber
&&-\frac{1}{6}\,\int\int \  \Big(\nabla{\tilde\Psi}({\#x},t;{\#x}_h,t_h)\Big) \times\*E({\#x}-{\#x}_h,t-t_h) \,
d{\#x}_h\, dt_h
\\
\nonumber
&&-\frac{1}{6}\,\int\int \  \Big(\partialt{\tilde\Psi}({\#x},t;{\#x}_h,t_h)\Big) \.\*B({\#x}-{\#x}_h,t-t_h) \,
d{\#x}_h\, dt_h
\\
&&\qquad\qquad\quad= 
\tilde{\#J}_{so}\rt\,.
\label{eq6.18a}
\end{eqnarray}

The second integral on the left side of \r{eq6.17a} is null--valued by virtue
of \r{eq6.19}; likewise, the third integral on the left side of \r{eq6.18a} is null--valued
by virtue of \r{eq6.20}. Therefore, the four macroscopic Maxwell postulates now read as follows:
\begin{eqnarray}
\nonumber
&&
\int\int \  \div\Big(\+\eps({\#x},t;{\#x}_h,t_h)\.\*E({\#x}-{\#x}_h,t-t_h)
\\
\nonumber
&&\qquad\qquad+\+\alpha({\#x},t;{\#x}_h,t_h)\.\*B({\#x}-{\#x}_h,t-t_h)\Big) \,
d{\#x}_h\, dt_h
\\
\nonumber
&&+\frac{1}{6}\,\int\int \  \Big(\nabla{\tilde\Psi}({\#x},t;{\#x}_h,t_h)\Big) \.\*B({\#x}-{\#x}_h,t-t_h) \,
d{\#x}_h\, dt_h
\\
&&\qquad\qquad\quad= \tilde{\rho}_{so}\rt\,,
\label{eq6.17b}
\\
\nonumber
&&
\int\int \  \curl\Big(\+\beta({\#x},t;{\#x}_h,t_h)\.\*E({\#x}-{\#x}_h,t-t_h)
\\
\nonumber
&&\qquad\qquad+\+\nu({\#x},t;{\#x}_h,t_h)\.\*B({\#x}-{\#x}_h,t-t_h)\Big) \,
d{\#x}_h\, dt_h
\\
\nonumber
&&
-\int\int \  \partialt\Big(\+\eps({\#x},t;{\#x}_h,t_h)\.\*E({\#x}-{\#x}_h,t-t_h)
\\
\nonumber
&&\qquad\qquad+\+\alpha({\#x},t;{\#x}_h,t_h)\.\*B({\#x}-{\#x}_h,t-t_h)\Big) \,
d{\#x}_h\, dt_h
\\
\nonumber
&&-\frac{1}{6}\,\int\int \  \Big(\nabla{\tilde\Psi}({\#x},t;{\#x}_h,t_h)\Big) \times\*E({\#x}-{\#x}_h,t-t_h) \,
d{\#x}_h\, dt_h
\\
\nonumber
&&-\frac{1}{6}\,\int\int \  \Big(\partialt{\tilde\Psi}({\#x},t;{\#x}_h,t_h)\Big) \.\*B({\#x}-{\#x}_h,t-t_h) \,
d{\#x}_h\, dt_h
\\
&&\qquad\qquad\quad= 
\tilde{\#J}_{so}\rt\,,
\label{eq6.18b}
\\
\label{eq6.19b}
&&\div\tilde{\#B}\rt = 0\,,
\\
\label{eq6.20b}
&&
\curl\tilde{\#E}\rt +
\partialt\tilde{\#B}\rt = \#0\,.
\end{eqnarray}

Differentiation of the product of two functions is distributive. Hence,
the  thirty--five independent constitutive
scalars in $\+\eps$, $\+\alpha$, $\+\beta$ and $\+\nu$ occur in  \r{eq6.17b}--\r{eq6.20b} 
in two ways: (i) by themselves, and (ii) through their space-- and time--derivatives. In
contrast, the thirty--sixth constitutive scalar ${\tilde\Psi}$ does not occur in  \r{eq6.17b}--\r{eq6.20b} by itself. 
Thus, $\tilde\Psi$ vanished from the macroscopic Maxwell postulates like the Cheshire cat, but left behind its derivatives 
like the cat's grin.

This is an anomalous situation, and  its elimination leads to the Post constraint.

\subsection{Post's Conclusions}
In a seminal contribution on the covariant structure
of modern electromagnetism \c{Post}, Post made a distinction
between functional and structural fields.  Functional fields specify the state of a medium, and are exemplified
by $\*E$ and $\*B$. Structural fields, exemplified by the constitutive tensors, specify the properties
of the medium. Formulating the Lagrangian and examining its Eulerian derivative \cite[Eq. 5.31]{Post},
Post arrived at the conclusion that
\begin{equation}
\label{Post6.18}
{\tilde\Psi}({\#x},t;{\#x}_h,t_h)\equiv 0
\end{equation}
even for nonhomogeneous mediums \cite[p. 130]{Post}. Furthermore, he held that the space-- and time--derivatives
of ${\tilde\Psi}({\#x},t;{\#x}_h,t_h)$ are also identically zero, so that \cite[p. 129]{Post}
\begin{equation}
\label{Post6.19}
\left.\begin{array}{ll}
\nabla{\tilde\Psi}({\#x},t;{\#x}_h,t_h)\equiv 0\\[5pt]
\partialt{\tilde\Psi}({\#x},t;{\#x}_h,t_h)\equiv 0
\end{array}\right\}\,.
\end{equation}

Equations \r{Post6.18} and \r{Post6.19} may appear to be independent but are not, because the derivatives of
a constant function are zero. Equation \r{Post6.18} alone is
 called the {\em Post constraint}.

\subsection{Recognizable Existence of $\tilde\Psi$}
Whether $\tilde\Psi$ is identically null--valued or not is a moot point. The real issue is
whether it has a recognizable
existence or not. This stance was adopted  by Lakhtakia
and Weiglhofer \c{WL98}.

Let us recall that all matter is microscopic. Despite the convenience proffered by
continuum theories, those theories are merely approximations. 
Constitutive functions are macroscopic entities arising from the homogenization 
of assemblies of microscopic charge carriers, with free space serving as the reference medium \c{W03}.
In any small enough portion of spacetime that is homogenizable, the constitutive functions are
uniform. When such a portion will be interrogated for characterization, it will have to be embedded
in free space. Accordingly, the second integral on the left side of \r{eq6.17b} as well as the third as well as the fourth integrals
on the left side of \r{eq6.18b} would vanish during the interrogation for fields inside and outside that
piece.
Therefore, the principle of parsimony (attributed to a 14th century monk \c{Occ}) enjoins the acceptance of
\r{Post6.18}.

\subsection{Nature of the Post Constraint}

When linear mediums of increasing complexity are investigated, the nature of thePost constraint can appear to vary.
For instance, were investigation confined to isotropic mediums \c{Lsst}, the condition
${\tilde\Psi}\equiv 0$ can resemble a {\em reciprocity constraint}. But it is not, because it does not impose
any transpose--symmetry requirements on $\+\eps$, $\+\alpha$, $\+\beta$ and $\+\nu$ \cite[Eqs. 23]{Kong}.

Another possibility is to think that the Post constraint negates the generalized duality transformation \c{SSTS},
but actually it does not when it is globally applied at the microscopic level \cite[pp. 203--204]{LWem}. Finally,
the Post constraint is not a gauge transformation~---~i.e., a ${\tilde\Psi}$--independent field ${\tilde{\#A}}$ cannot be found to replace
${\tilde{\#P}}$ and
${\tilde{\#M}}$   by ${\tilde{\#P}}-\curl {\tilde{\#A}}$
and  ${\tilde{\#M}}+(\partial/\partial t)\,{\tilde{\#A}}$,
respectively, in order to eliminate ${\tilde\Psi}$.

The Post constraint is actually a {\em structural constraint}. Post may have been inspired
towards it in order to eliminate a pathological constitutive
relation \cite[Eq. 3.20]{Post}, \c{LWima}, and then established a covariance argument
for it. Physically, this constraint arises from the following two considerations:
\begin{itemize}
\item The Amp\`ere--Maxwell equation (containing the induction fields) should be independent
of the Faraday equation (containing the primitive fields) at the macroscopic level, just
as the two equations are mutually independent at the microscopic level.
\item The constitutive functions must be characterized as piecewise
uniform, being born of the spatial homogenization of microscopic entities. Therefore,
if a homogeneous piece of a medium with a certain set of electromagnetic response properties
cannot be recognized, the assumption of  continuously nonhomogeneous analogs of that set
is untenable.
\end{itemize}

\section{EH--Electromagnetism}
\label{EHEM}

Time--domain electromagnetic research is a distant second to frequency--domain electromagnetic
research, as measured by the numbers of publications as well as the numbers of researchers. Much of frequency--domain
research at the macroscopic level also commences with the familar form \r{eq6.17}--\r{eq6.20} of the Maxwell postulates,
but  the roles of $\*H$ and $\*B$ are interchanged \c{W03}. 

Thus, constitutive relations are written to express $\*D$ and $\*B$
in terms of $\*E$ and $\*H$. Specifically, the linear constitutive relations \r{dconrel1} and \r{hconrel1} are replaced by
\begin{eqnarray}
\nonumber
&&\*D\rt = \int\int \  \+{\cal A}({\#x},t;{\#x}_h,t_h)\.\*E({\#x}-{\#x}_h,t-t_h) \,
d{\#x}_h\, dt_h
\\[5pt]
\label{dconrel2}
& &\qquad+\int\int \  \+{\cal B}({\#x},t;{\#x}_h,t_h)\.\*H({\#x}-{\#x}_h,t-t_h) \,
d{\#x}_h \,dt_h \, 
\end{eqnarray}
and
\begin{eqnarray}
\nonumber
&&\*B\rt = \int\int \  \+{\cal C}({\#x},t;{\#x}_h,t_h)\.\*E({\#x}-{\#x}_h,t-t_h) \,
d{\#x}_h\, dt_h
\\[5pt]
\label{bconrel2}
&&\qquad+\int\int \  \+{\cal D}({\#x},t;{\#x}_h,t_h)\.\*H({\#x}-{\#x}_h,t-t_h) \,
d{\#x}_h\, dt_h \, ,
\end{eqnarray}
with $\+{\cal A}$, $\+{\cal B}$, $\+{\cal C}$ and $\+{\cal D}$ as the constitutive tensors.
This version of electromagnetism is called the EH--electromagnetism in this essay.

At first glance, the difference between the modern and the EH versions may not appear to be
significant, particularly for linear mediums at the macroscopic level. The frequency--domain versions
of the constitutive tensors $\+{\cal A}$, etc., can also be microscopically motivated in much the same
way as the frequency--domain versions of $\+\eps$, etc., are. Yet, there is a huge difference:
The Faraday equation contains only the primitive fields  while
the Amp\`ere--Maxwell equation contains only the induction fields, in modern electromagnetism, and can therefore be
independent of each other just as at the microscopic level.
But each of the two equations contains a primitive field and an induction field in EH--electromagnetism~---~hence,
it is impossible for the two equations to be independent of each other at the macroscopic level.
This central difference between the two versions of electromagnetism
is often  a source of great confusion.

\subsection{Post Constraint}

As both the Faraday and the Amp\`ere--Maxwell equations (at the macroscopic
level) contain a primitive field and an induction field
in EH--electromagnetism, it appears impossible to derive the Post constraint in the EH version. Not surprisingly,
current opposition to the validity of the
Post constraint invariably employs the EH version \c{SSTS,TMNVBS}, and older constructs that presumably
invalidate the Post constraint are also based on EH--electromagnetism \c{Tell,Dzya,FRS}. The major
exception to the previous statement is the work of O'Dell \cite[pp. 38--44]{ODell}, but it is fatally marred by the assumption
of purely instantaneous~---~and, therefore, noncausal~---~constitutive relations. Simply put, {\em the Post constraint is valid
in modern electromagnetism but probably invalid in EH--electromagnetism}.

But we hold modern electromagnetism to be {\em truer} than its EH counterpart \c{Jack, Sch, LDZ,Post03}. Accordingly,
the Post constraint can {\em translated} from the former to the latter, in certain circumstances. For example, let us
consider a spatially homogeneous, temporally invariant and spatially local medium: $\+\eps({\#x},t;{\#x}_h,t_h) \equiv \+\eps(t_h)\,\delta({\#x}_h)$, etc.
Employing the temporal Fourier transform\footnote{Whereas all quantities decorated with a tilde $\tilde{}$ are real--valued, their
undecorated counterparts are complex--valued in general.}
\begin{equation}
{\tilde Z}\rt =\frac{1}{2\pi}\int_{-\infty}^\infty\, Z\ro \, \exp(-i\omega t) \,d\omega\,,
\end{equation}
where $\omega$ is the angular frequency and $i=\sqrt{-1}$,
we see that \r{dconrel1} and \r{hconrel1} transform to
\begin{equation}
\left.\begin{array}{ll}
\#D\ro = \=\eps(\omega)\.\#E\ro + \=\xi(\omega)\.\#B\ro\\
\#H\ro = \=\zeta(\omega)\.\#E\ro + \=\nu(\omega)\.\#B\ro
\end{array}\right\},
\end{equation}
while \r{dconrel2} and \r{bconrel2} yield
\begin{equation}
\left.\begin{array}{ll}
\#D\ro = \={\cal A}(\omega)\.\#E\ro + \={\cal B}(\omega)\.\#H\ro\\
\#B\ro = \={\cal C}(\omega)\.\#E\ro + \={\cal D}(\omega)\.\#H\ro
\end{array}\right\}.
\end{equation}
With the assumption that $\={\cal D}(\omega)$ is invertible, the Post constraint
\begin{equation}
\Psi(\omega) \equiv 0
\label{pqm}
\end{equation}
translates into the condition \c{WL94}
\begin{equation}
{\rm Trace}\Big(\={\cal B}(\omega)\.\={\cal D}^{-1}(\omega)+ \={\cal D}^{-1}(\omega)\.\={\cal C}(\omega)\Big)\equiv0
\label{pqEH0}
\end{equation}
for EH--electromagnetism; equivalently,
\begin{equation}
{\rm Trace}\Big[\={\cal D}^{-1}(\omega)\. \Big(\={\cal B}(\omega)+\={\cal C}(\omega)\Big)\Big]\equiv0\,.
\label{pqEH}
\end{equation}
We must remember, however, that \r{pqEH} is probably underivable within the framework
of EH--electromagnetism, but is simply a translation of \r{pqm}.

\section{Experimental Evidence}
\label{ExpEvi}
Fundamental questions are answered by a convergence of theoretical constructs and diverse
experimental evidence. On this basis, modern electromagnetism is well--established, which provides
confidence in the validity of the Post constraint. Furthermore, incontrovertible experimental
results against the Post constraint are unknown. Nevertheless, the constraint is experimentally
falsifiable, and available experimental evidence presented against
it must not be dismissed lightly. Let us examine that evidence now.

\subsection{Magnetoelectric Materials}
\label{MEmat}
Anisotropic materials with magnetoelectric tensors are commonplace. Typically, such
properties are exhibited at low frequencies and low temperatures. Although their emergence in
research literature can be traced back to Pierre Curie \c{Schmid03}, a paper
published originally in 1959 \c{Dzya} focused attention on them. O'Dell wrote a famous book
on these materials \c{ODell} in 1970.

A significant result of O'Dell \cite[Eq. 2.64]{ODell}, although derived for spatiotemporally uniform and spatiotemporally
local mediums (i.e., $\+\eps(\#x,t;{\#x}_h,t_h) = \+\eps\, \delta({\#x}_h)\,\delta(t_h)$, etc.), is often used in frequency--domain
literature on spatiotemporally uniform and spatially local mediums as follows:
\begin{equation}
\label{OD}
{\rm Transpose}\Big(\=\xi(\omega)\Big)=- \=\zeta(\omega)\,.
\end{equation}
This equation is often held to allow materials for which $\Psi(\omega)\ne 0$. More
importantly, this equation is widely used in the magnetoelectric
research community to reduce experimental tedium in characterizing magnetoelectric materials. Yet this equation is based
on a false premise: that materials (as distinct from free space) respond purely instantaneously \cite[p. 43]{ODell}.
Hence, experimental data obtained after exploiting \r{OD} cannot be trusted \c{Linv}.

The false premise can be traced back to Dzyaloshinski\u{i}'s 1959 paper \c{Dzya}, wherein EH--electromagnetism
was used. Astrov \c{Astrov} examined the variation of $\={\cal C}(\omega)$ of Cr$_2$O$_3$  with temperature at 10~kHz frequency.
Folen {\em et al.\/} \c{FRS61} measured $\={\cal C}(\omega)$ of Cr$_2$O$_3$ at 1~kHz frequency and
presumably equated it to $\={\cal B}(\omega)$
by virtue of the 1959 antecedent \c{Rado}, but did not actually measure $\={\cal B}(\omega)$.\footnote{This and a large fraction of
other published reports do not seem to recognize that ${\underline{\underline{\cal C} }}(\omega)$, etc., are complex--valued quantities, but treat them
as real--valued quantities.}
Rado and Folen
\c{RF1,RF2} verified the existences of both $\={\cal B}(\omega)$ and $\={\cal C}(\omega)$ for the same
substance, and they also established that both quantities are temperature--dependent, but they too did
not measure $\={\cal B}(\omega)$.
Similar deficiencies in
other published reports have been detailed elsewhere \c{Linv}.  Recently, Raab \c{R03} has rightly called for comprehensive and
complete characterization of magnetoelectric materials, with \r{OD} not assumed in advance but actually subjected to a test.

\subsection{Tellegen Medium}
\label{TellMed}
Take a fluid medium in which permanent, orientable, electric dipoles exist in abundance. 
Stir in small ferromagnetic particles with permanent magnetic dipole moments, 
ensuring that each electric dipole moment cleaves together with a parallel magnetic dipole moment,
to form a Tellegen particle \c{TMNVBS}.
Shake well for a homogeneous, isotropic suspension of Tellegen particles. This is the recipe that Tellegen
\c{Tell} gave for the so--called Tellegen medium, after he had conceptualized the gyrator.

The frequency--domain constitutive relations of this medium
may be set down as 
\begin{equation}
\label{Tell1}
\left.\begin{array}{ll}
\#D\ro = {\cal A}(\omega)\,\#E\ro + {\cal B}(\omega)\,\#H\ro\\
\#B\ro = {\cal B}(\omega)\,\#E\ro + {\cal D}(\omega)\,\#H\ro
\end{array}\right\}\,,
\end{equation}
with the assumption of temporal invariance, spatial homogeneity, spatial locality, and isotropy.
Furthermore, \r{Tell1} apply only at sufficiently low frequencies \c{LOCM}.

Gyrators have been approximately realized using other circuit elements, 
but the Tellegen medium has never been successfully synthesized. Tellegen's own experiments
failed \cite[p. 96]{Tell}
Neither has the Tellegen medium been observed in nature. Hence, non--zero values
of ${\cal B}(\omega)$ of actual materials are not known. A fairly elementary exercise shows that the
recognizable existence of this medium is tied to that of irreducible magnetic sources \c{Booj,Dmit}.
As the prospects of observing a magnetic monopole are rather remote \c{JL95,Hagi}, for now it is appropriate to regard the
Tellegen medium as chimerical.

\subsection{Tellegen Particle}
\label{TellP}
Each particle in Tellegen's recipe is actually a uniaxial particle \c{WLM93}. Because the recipe calls for
the suspension to be homogeneous, the particles cannot be similarly oriented. However, if all particles were
similarly oriented in free space, and the number density $N_p$ of the particles is very small, the frequency--domain
constitutive relations of the suspension at sufficiently low frequencies will be
\begin{equation}
\left.\begin{array}{ll}
\#D\ro \simeq \epso\#E\ro +N_p\Big(\={\pi}^{(ee)}(\omega)\.\#E\ro + \={\pi}^{(eh)}(\omega)\.\#H\ro\Big)\\[5pt]
\#B\ro \simeq \muo\#H\ro+ N_p\Big(\={\pi}^{(he)}(\omega)\.\#E\ro + \={\pi}^{(hh)}(\omega)\.\#H\ro\Big)
\end{array}\right\}\,,
\end{equation}
wherein $\={\pi}^{(ee)}$, etc., are the polarizability tensors of a Tellegen particle in free space.

A recent report \c{TMNVBS} provides experimental evidence on the existence of ${\=\pi}^{(eh)}$ for a Tellegen
particle made by sticking a short copper wire to a ferrite sphere biased by a quasistatic magnetic field parallel
to the wire. However, this work can not lead to any significant finding against the
validity of the Post constraint  for the following two reasons:
\begin{itemize}
\item Although a quantity proportional to the magnitude of ${\rm Trace}\big({\=\pi}^{(eh)}\big)$ was measured, a similar measurement of
${\rm Trace}\big({\=\pi}^{(he)}\big)$ was not undertaken; instead, the identity  
\begin{equation}
{\rm Trace}\big({\=\pi}^{(he)}(\omega)\big)=  {\rm Trace}\big({\=\pi}^{(eh)}(\omega)\big)
\end{equation} 
was assumed without testing. This deficiency in experimentation is similar to that for magnetoelectric materials
mentioned in Section \ref{MEmat}.

\item The Post constraint is supposed to hold rigorously for linear electromagnetic response with respect to
the total electromagnetic field, which is constituted jointly by
the bias magnetic field as well as the time--harmonic electromagnetic
field. As discussed by Chen \c{Chen}, the ferrite is therefore a nonlinear material.

\end{itemize}
Incidentally, the biased--ferrite--metal--wire modality for Tellegen particles is likely to be very difficult to implement
to realize the Tellegen medium of Section \ref{TellMed}.

\subsection{Summation of Experimental Evidence}
On reviewing Sections \ref{MEmat}--\ref{TellP}, it becomes clear that experimental evidence against the validity
of the Post constraint is incomplete and inconclusive, in addition to being based either on the false
premise of purely instantaneous response and/or derived from EH--electromagnetism.

\section{Post Constraint and Free Space}
\label{PCfs}
Although the Post constraint holds for modern electromagnetism, which has a microscopic basis
in that matter is viewed as an assembly of charge--carriers in free space, before concluding this essay
it is instructive to derive
the constitutive equations of free space back from the macroscopic constitutive equations \r{dconrel1} and \r{hconrel1}.

Let us begin with free space being spatio\-temporally invariant and spatio\-temporally local; then,
$\+\eps({\#x},t;{\#x}_h,t_h) = {\+\eps}_{fs}\, \delta({\#x}_h)\,\delta(t_h)$, etc., and \r{dconrel1} and \r{hconrel1}
simplify to
\begin{equation}
\left.\begin{array}{ll}
\*D\rt = \+\eps_{fs}\.\*E\rt + \+\xi_{fs}\.\*B\rt \\
\*H\rt = \+\zeta_{fs}\.\*E\rt + \+\nu_{fs}\.\*B\rt 
\end{array}\right\}\,.
\label{conrelfs1}
\end{equation}
The free energy being a perfect differential, and because the constitutive relations
\r{conrelfs1} do not involve convolution integrals, it follows that \cite[Eq. 6.14]{Post}
\begin{equation}
{\rm Transpose}\big(\+\xi_{fs}\big)=- \+\zeta_{fs}\,.
\label{treq}
\end{equation}
With the additional requirement of isotropy, we get
\begin{equation}
\left.\begin{array}{ll}
\*D\rt ={\tilde \eps}_{fs}\,\*E\rt +{\tilde \xi}_{fs}\,\*B\rt \\
\*H\rt = -{\tilde\xi}_{fs}\,\*E\rt + {\tilde\nu}_{fs}\,\*B\rt 
\end{array}\right\}\,.
\label{conrelfs2}
\end{equation}
The subsequent imposition of the Post constraint means that $\xi_{fs}=0$, and the
constitutive relations
\begin{equation}
\left.\begin{array}{ll}
\*D\rt = {\tilde\eps}_{fs}\,\*E\rt  \\
\*H\rt =  {\tilde\nu}_{fs}\,\*B\rt 
\end{array}\right\}\,.
\label{conrelfs3}
\end{equation}
finally emerge.  The values ${\tilde\eps}_{fs}=\epso$ and ${\tilde\nu}_{fs} = 1/\muo$ are used
in SI \c{Post03}.
Although Lorentz--reciprocity was not explicitly enforced for free space, it emerges naturally in this 
exercise \c{WLjosa}. Alternatively, it could have been enforced from the very beginning, and it would
have led to  ${\tilde\xi}_{fs}=0$ \c{BPS81}. 

\section{Concluding Remarks}
Despite the fact that the mathematical forms of the macroscopic Maxwell postulates are identical
in modern electromagnetism as well as in EH--electromagnetism, the two are very physically very different.
Modern electromagnetism is held to be basic; hence, the answers to all fundamental questions
must be decided within its framework. Thereafter, if necessary, its equations can be transformed into the
frequency domain and {\em then\/} into those  of EH--electromagnetism~---~and the resulting equations may be used
to solve any  problems that a researcher may be interested in. The reverse transition from EH--electromagnetism
to modern electromagnetism can lead to false propositions.

\section*{Acknowledgment} Occasional discussions with Dr. E.J. Post are gratefully acknowledged.


\end{document}